\documentstyle{article}

\begin{document}
\renewcommand{\theequation}{\arabic{section}.\arabic{equation}}
\setcounter{equation}{0}
\setcounter{section}{0}

\title{Non  Abelian Dual Maps in Path Space} 
\author{Isbelia  Mart\'{\i}n \\
{\it  Depto. de F\'{\i}sica, Universidad Simon Bol\'{\i}var}\\
{\it Caracas 89000, Venezuela. e-mail: isbeliam@usb.ve}}
\maketitle  
\begin{abstract} We study an extension of the procedure to construct duality
transformations among abelian gauge theories to the non abelian case using a path space
formulation. We define a pre-dual functional in path space and introduce a particular non local map among Lie 
algebra valued 1-form functionals that reduces to
the ordinary Hodge-${\star}$ duality map of the abelian theories. Further, we 
establish a full set of equations on path space representing the ordinary 
Yang Mills equations and Bianchi identities of non abelian gauge 
theories of 4-dimensional euclidean space. 
\end{abstract}


\section{Introduction}
Duality maps between abelian gauge theories given by $U(1)$ connections on line bundles over a
manifold $X$ can be shown to exist  by using a  quantum equivalent formulation of the original
theory in terms of closed 2-forms. This formulation was first shown to exist for abelian and non
abelian potential theories in a particular gauge
by Halpern \cite{Hal} and used, afterwards, for non abelian theories in some other gauge by
\cite{Durand}. The construction of quantum equivalent dual abelian theories 
was successfully achieved for trivial bundles in \cite{Kazama} using a 
kind of Fourier transform to the space of dual Hodge-${\star}$ forms. In 
the same article, they extend the construction
to the case of non abelian gauge theories but arrived to a dual theory with a parity violating
action arising from the ambiguity in the solution to a Bianchi type constraint.

Recently, the quantum equivalent dual abelian gauge theory was formulated for non trivial
bundles. It is achieved by defining first a functional on the space of abelian
2-forms which must be  constrained by non-local restrictions, namely, the requeriment of 
being closed and with integral periods, ensuring the existence of a 1-1 correspondence between the space of 
constrained 2-forms and the line bundles over $X$ \cite{Alv}.This procedure has
been successfully applied even to more general U(1) bundles \cite{Alv2} based on an
extension of Weil's theorem to complex p-forms \cite{Bry}. Once the equivalence between the
formulation in terms of the configuration space of abelian connections and that of the space
of closed 2-forms is achieved, the latter is used to construct at the quantum level the dual
gauge theory, by introducing dual Hodge-${\star}$ forms through Lagrange 
multipliers much in the spirit of \cite{Kazama}, proving 
the existence of non trivial relations between the partition functions of the abelian theory
and its dual.

The purpose of this article is to extend the above procedure  
\cite{Alv},\cite{Alv2} to the non abelian case without the ambiguities 
mentioned earlier. In the first place, we will begin by asking what conditions should be
imposed on matrix-valued 2-forms over a manifold $X$ to produce something
similar to Weil's theorem for non-abelian 2-forms, so that we could achieve an equivalence
between the formulation of the theory on the configuration space of connections and the
formulation on the space of 2-forms. This actually is a formidable problem still not solved
but only to the level of conjectures \cite{Bry2}. In any case, we could try to see where
failures lie and propose solutions. \cite{Isb} 

The Bianchi identity for a matrix-valued 2-form $\Omega $ \[ {\cal D}\;\Omega =
0 \] is the first condition that comes to mind when looking for restrictions to implement, since
curvatures for connections on fiber bundles satisfy it. But this, in general, does not assure even that
 $\Omega$ may be expressed locally in terms of any 1-form connection $A$ as \[ \Omega = dA + A\wedge A \] something
equivalent to a Poincar\'e's lemma for "covariantly closed" forms does not hold. Moreover, even
when we could express $\Omega$ in terms of $A$ as above, on open sets $U_i$ of a covering of the manifold
$X$, compatibility of the curvature-like 2-forms $\Omega(A_i)$ and $\Omega(A_j)$ on the
intersection of two open sets $U_i$ and $U_j$ should imply that $A_i$ and $A_j$ are related by
a well defined gauge transformation on the intersection of open sets. Simple calculations show
that this is not the case. $A_i$ and $A_j$ could be related by some other more general
transformations that no doubt include the mentioned gauge transformations i.e. \[ \Omega (A_i)
= g^{-1} \Omega(A_j)\; g \;  \stackrel{not}{\Longrightarrow}\; A_i = g^{-1}A_j \;g + g^{-1}dg
\]
this is exactly the root of the ambiguity problem in \cite{Kazama}.  

Obviously, we need more restrictive conditions to arrive to the necessary compatibility
glueing for constructing globally well defined non abelian principal bundles. 
In \cite{Durand} this restrictive condition was  partially found but 
at the expense of introducing disguised path dependent gauge 
transformations.

On the other hand, it is well known \cite{Poly}, that a formulation of non abelian gauge theories has a rather
simple expression on the space of loops as a trivial flat gauge theory. The main ingredient in
this formulation is the use of the holonomy associated to each class of  non abelian Lie
algebra valued connections on a principal bundle. The use of holonomies  is quite adequate since
its non local character as a geometrical object carries a lot more information about the
bundle than curvatures or connections do. So, we should go to the loop space formulation and see
whether it is possible to write some conditions that could characterize the non abelian
bundles and look for a procedure to build the duality maps. In what follows, we suceed in
proving half the task, for a more detailed discussion see \cite{LIA}.

For our purpose, instead of using the space of closed curves \cite{Poly,Barre}, we will consider
a space of open curves ${\cal C}$  \cite{Bry2} with fixed endpoints {\rm O}, {\rm P} 
over a smooth connected 4-dimensional euclidean manifold $X$ with reference point {\rm O} .
This will allow the construction of  smoothly behaving mathematical objects like
functionals, variations of functionals, 1-form connection functionals and so on, on open
neighborhoods of the space of curves. Particularly, we avoid regularization problems in the
definition of the gauge "potential" on loop space.

\section{ Connections and curvatures in path space} 

First, any functional over this space will be denoted $ \widetilde {\Phi} ({\cal
C}_{\;\rm{O,P}}) $ and a variation or increment of this functional due 
to a small deformation on the curve leaving the endpoints fixed is defined as
 \[ \widetilde {\Delta} \widetilde {\Phi}
({\cal C}_{\;\rm{O,P}}) \equiv \widetilde {\Phi} ({\cal C}_{\;\rm {O,P}}+\delta {\cal
C}_{\;\rm {O,P}}) - \widetilde {\Phi} ({\cal C}_{\;\rm{O,P}}) \] 
Deformations on the curves are smooth vector fields on open neighborhoods of $X$ where the curve ${\cal C}_{\;\rm{O,P}}$
lies,  tending to zero on the endpoints of the curve. We could relax this definition allowing
non zero deformations on one of the endpoints but then  we would need to impose a non linear
condition to get the compatibility requirement on the patching of the principal bundle
\cite{LIA}.

Our version of ``holonomy'' is $G_A$, the path ordered exponential of a 1-form
connection $A$ over $X$ integrated over the open curve ${\cal{C}}$ i.e 
\[  G_{A} ({\rm O},{\rm P},{\cal
C}) \equiv exp :\int_{\rm O}^{\rm P}A : \]
it becomes the ordinary holonomy when {\rm O} and {\rm P} are 
identified. It is a map from $PX$ to the non abelian group ${\cal{G}}$.

$\widetilde {\cal A}({\cal {C}}_{\;\rm{O,P}})$ denotes the Lie algebra 
valued 1-form `connection' functional acting on
small deformations $S$.
and may be expressed in terms of $ F_{p(t')}$ , the ordinary pointwise defined curvature 2-form
associated to the connection $A$, as
\begin{equation} \widetilde {\cal A}({\cal C}_{\;\rm{O,P}}) [S]
= \int_{\rm O}^{\rm P} G_{A} ({\rm O},p(t'),{\cal C}) F_{\mu \nu}(t') 
T^{\mu}(t')S^{\nu}(t') 
G_{A} ({\rm
O},p(t'),{\cal C}^{-1}) dt' \label{1} \end{equation}
where $T$ is a vector field tangent to the curve ${\cal C}$, $t'$ is a parameter along the curve
and $ p(t')$ is an ordinary point on the curve. In general, it may be 
written as an object acting on any small deformation as
\[ \widetilde {\cal A}({\cal C}_{\;\rm{O,P}})[\bullet] 
= \int_{\rm O}^{\rm P} G_{A} ({\rm O},p(t'),{\cal C}) F(t') [T,\bullet] 
G_{A} ({\rm
O},p(t'),{\cal C}^{-1}) dt' \] 
It is obtained from $G_A$
\begin{eqnarray}
\widetilde{A}({\cal C}_{\;\rm{O,P}}) [S] &=&G(P,{\cal C})(A(%
\Delta P))G^{-1}(P,{\cal C})-  \nonumber \\
&&-\tilde{\Delta}G(P,{\cal C})[S] G^{-1}(P,%
{\cal C}).
\end{eqnarray}
on small deformations $S$ of the curve ${\cal {C}}$ with non zero 
variation of the endpoint $\rm P$. Notice that
\begin{eqnarray}
G(P+\Delta P,{\cal C}_{S})G^{-1}(P,{\cal C}) &=&G(P+S_{P}\Delta s,{\cal C}_{s})({\rm 1%
\hspace{-0.9mm}l\hspace{-1mm}}-(A[S_{P}]\Delta s)G^{-1}(P,{\cal C})+  \nonumber
\\
&&+G(P+S_{P}\Delta s,{\cal C}_{S})(A[S_{P}]\Delta s) G^{-1}(P,{\cal C})+G(P+S_{P}\Delta s,{\cal C}%
_{S}).  \nonumber \\
.(A[T_{P}]\Delta t) G^{-1}(P,{\cal C}) &=&G(P+S_{P}\Delta s,{\cal C}_{S})({\rm 1\hspace{%
-0.9mm}l\hspace{-1mm}}-(A[S_{P}]\Delta s)G^{-1}(P,{\cal C}) \nonumber \\
&&+G(P,{\cal C})(A(\Delta P))G^{-1}(P,{\cal C})
\end{eqnarray}
where ${\cal C}_{S}$ is short for ${\cal C}_{\;\rm {O,P}}+\delta {\cal
C}_{\;\rm {O,P}}$. The last equality holds to first order in ${\Delta}s$ 
and ${\Delta}t$,and 
\[
\Delta P=S_{P}\Delta s + T_{P}\Delta t.
\]
We then obtain to first order in $\Delta s$ 
\begin{eqnarray}
\widetilde{A}({\cal C}_{\;\rm{O,P}}) [S]\Delta s &=&-G(P+S_{P}\Delta s,{\cal C}_{S})({\rm 1\hspace{%
-0.9mm}l\hspace{-1mm}}-(A[S_{P}]\Delta s)G^{-1}(P,{\cal C}) 
\nonumber \\
&&+{\rm 1\hspace{-0.9mm}l\hspace{-1mm}}  \nonumber \\
&=&-\exp :\int_{0}^{t_{P}}-G(t,{\cal C})F_{\mu \omega }(t){S^{\omega }}%
(t) G^{-1}(t,{\cal C})dt \Delta s:+{\rm 1\hspace{-0.9mm}l\hspace{-1mm}} 
\nonumber \\
&=&\int_{0}^{t_{P}}G(t,{\cal C})F_{\mu \omega }(t)T^{\mu }(t)S^{\omega
}(t)G^{-1}(t,{\cal C})dt \Delta s
\end{eqnarray}
that is eq. \ref{1}.

For zero deformations on the endpoint $\rm P$ then
\begin{equation}
\widetilde{A}({\cal C}_{\;\rm{O,P}})  =
-\tilde{\Delta}G(P,{\cal C})\cdot G^{-1}(P,%
{\cal C}).
\label{2}¥\end{equation}
and eq.\ref{1} still holds.
 
$ \widetilde {\cal A}({\cal {C}})$ is defined for classes of equivalence of ordinary connections
under gauge transformations, i.e it is gauge invariant up to elements of the structure group on
the endpoints of the curve. This is a rather nice feature of working in 
path or loop spaces . Also, it depends on the curve  ${\cal {C}}$. 
From now on we will denote $\widetilde {\cal A}({\cal C}_{\;\rm{O,P}})$ as $\widetilde {\cal 
A}(P,{\cal {C}})$ making reference only to the endpoint $\rm P$. 

We could continue and define also the Lie algebra valued `curvature' functional $ \widetilde 
{\cal {F}} ({\cal
C},A)$ for the `connection' functional $ \widetilde {\cal A}$ in the usual manner

\[ \widetilde {\cal F} ({\cal {C}},A) = \widetilde {\Delta} \widetilde {\cal A} ({\cal C})
 \; + \; \widetilde {\cal A} ({\cal C}) \wedge \widetilde {\cal A} ({\cal C})  \]
 for this free formulation of non abelian gauge theories, one shows 
 easily that for any small deformation with fixed endpoints
\[\widetilde {\cal F} ({\cal C},A) \;=\; 0 \]
and it is a gauge invariant statement. See that eq.\ref{2}  implies that 
\begin{eqnarray}
\widetilde{\Delta}(\widetilde{\Delta}G) &=&-\widetilde{\Delta}(\widetilde{A}G)=(\widetilde{\Delta%
}\widetilde{A}\cdot G-\widetilde{A}\wedge \widetilde{\Delta}G)  \nonumber \\
&=&(\widetilde{\Delta}\widetilde{A}+\widetilde{A}\wedge \widetilde{A})G=0.
\label{3}¥\end{eqnarray}
this equation \ref{3} shows an interesting property of 
$\widetilde{\Delta}$, namely
\[ \widetilde{\Delta}\widetilde{\Delta}\bullet=0\] reminiscent of an 
exterior differential operator.
 
The "covariant" derivative $\widetilde{\cal D}$ may be also introduced as
\[ \widetilde{\cal D} \;\cdot \;\; \equiv \;\; \widetilde {\Delta}\;\cdot\;\; +\; \widetilde
{\cal A} \wedge\; \cdot \]

Now on the space of curves and deformations with  fixed endpoints  \cite{LIA},
we may impose a flatness condition on general 1-forms functionals $\widetilde{\omega}$ that can lead only 
to locally well defined non abelian principal bundles. 

Let us see how it works, take a flat 1-form functional $\tilde{\omega}$ on $PX$ . We would like to know
about the existence of a non abelian bundle and of 1-form connections $%
A$ on $X$ for which 
\begin{equation}
\widetilde {\omega}(P,{\cal C})=\widetilde {A}(P,{\cal C}).
\end{equation}
in here
\begin{equation}
\widetilde{\omega}(P,{\cal C})=\int_{0}^{t_{P}}G(t,{\cal C})\theta _{\mu
\omega }(t)T^{\mu }(t)d{X_{t}}^{\omega }{G}^{-1}(t,{\cal C})dt
\end{equation}
where $\theta _{\mu \upsilon }$ is a smooth general antisymmetric field 
on $X$ and $G(t,{\cal C})$ is the path ordered exponential of an 
arbitrary 1-form connection $A$ and we have taken
\[\theta [T,\bullet] = \theta _{\mu\omega }(t)T^{\mu }(t)d{X_{t}}^{\omega 
}[\bullet] \] in some particular local coordinate system $\{X^{\mu}¥\}¥$. 

The flatness condition reads,
\begin{equation}
\widetilde{\Delta}\widetilde{\omega }+\widetilde{\omega }\wedge \widetilde{%
\omega }=0,
\label{7}¥\end{equation}

Consider an open covering of $X$ and on $U_{1}$ we solve the flatness
condition for $\tilde{\omega}$  obtaining 
\begin{equation}
\widetilde{\omega }(P,{\cal C})=\int_{{\cal C}}^{t_{P}}GF_{\mu \omega
}T^{\mu }d{X_{t}}^{\omega }G^{-1}dt=\widetilde{A}(P,{\cal C}).
\end{equation}
so the unique  solution, up to gauge transformations, is after some 
lenghty but simple calculations
\[ \theta _{\mu \omega }(t) = F_{\mu \omega}(t)\]
with $ F_{\mu \omega}(t)$ the Lie algebra valued curvature field with 
connection $A$, see appendix B of  \cite{LIA} for details. 

By solving the same condition on $U_{2}$ we obtain 
\begin{equation}
\widetilde {\omega}(P,{\cal C})=\widetilde {A^{\prime}}(P,{\cal C})
\end{equation}
in terms of another local pointwise 1-form connection $A^{\prime}$.

On the intersection $U_{1}\cap U_{2}\neq \phi$ we have, up to gauge 
transformations, 
\begin{equation}
\widetilde {A}(P,{\cal C})= \widetilde {A^{\prime}}(P,{\cal C})
\end{equation}
and after some calculations we then get 
\begin{equation}
A^{\prime}=g^{-1}Ag + g^{-1}dg.
\end{equation}

i.e. the only way of a consistent patching  it is through a matching of
1-form functionals on intersections of $PX$,  corresponding to gauge 
transformations between pointwise 1-form connections on the 
intersection of neighborhoods 
$U_{1}$ and $U_{2}$ of $X$ . We have thus avoided 
the obstruction discussed earlier by working on $PX$ with the generalized 
1-form functional $\tilde{A}$.

The flatness condition does not assure that $A$ is a 1-form connection 
on a non abelian principal bundle over $X$. As in the abelian case, a global restriction is needed
to ensure that local 1-forms with the transition defined through 
gauge transformations, give rise to a globally defined 1-form connection on $X$.  A characterization of
matrix-valued 2-forms for being curvatures of non abelian bundles has also recently been
conjectured using partial differential equations on a loop space \cite{Bry2}.
 
Thus, we have succeeded in eliminating the local obstruction to the construction of dual maps. 
Now, it rests to find a global condition equivalent to that of integral periods of the
curvature 2-form for abelian gauge theories, that actually labels the different line bundles,
i.e an equivalent Dirac quantization condition. We know that for particular $SU(2)$ bundles, 
there may be a splitting  into the direct sum of two line bundles, for those bundles the usual Dirac
quantization may suffice. In the space $PX$ then the restriction to be imposed would be that
the ordinary curvature 2-form appearing in $ \widetilde {\cal A}({\cal C}_{\;\rm{O,P}})$ would
belong to the set of "diagonalizable" 2-form curvatures through a condition involving the
intersection form and the non abelian topological charge associated to the second Chern class.
This suggests that perhaps the global condition needed for non abelian gauge theories, at least
for the case of $SU(N)$, involves the "quantization" of the topological charge  associated to
the second Chern class. Once we find the exact global condition to be imposed 
on $PX$, and the dual map for constructing dual gauge theories  realized 
through an operation on $PX$ that generalizes the Hodge-$\star$ 
operation, we could arrive to an algorithm to get quantum equivalent 
non abelian dual theories.

\section{ Dual maps: the pre-dual 1-form functional $\tilde{B}$} 

In what follows we concentrate on defining dual maps on $PX$.  Having shown that 
the formulation of Yang - Mills on $PX$ avoids the problem
of the construction of a sequence for dual transformations in terms of
curvatures, we will now discuss the construction of the generalization of
the  Hodge-$\star$ operation and then show that the Yang - Mills field equations
can be rewritten in a form which are manifestly invariant under dual
transformations. An attempt to define that operation for loop spaces has  
been given in \cite{Cho}. In our construction we avoid
infinity factors which appear in their work, as well as several
regularizations in their calculations.  The regularizations are avoided by
working directly with the exterior algebra on $PX$ as defined in \cite{LIA}.

We introduce a Lie algebra valued 1-form functional over $PX$, $\tilde{B}$ the pre-dual 1-form 
functional, constructed from the Hodge-$\star$ dual of the curvature 2-form. 

First we construct a 2-surface $\Sigma$ in the following way. Given ${\cal {C}}\in PX$ 
we consider a family of curves ${\hat{\cal{C}}}_t$
obtained by a closed finite but small deformation that starts  and ends at
${\cal {C}}$  with tangent vectors $S$ and parametrized by $\sigma$. We 
identify $\rm t$ with the parametrization of the family of curves going smoothly 
from ${\rm O}$ to ${\rm P}$ with tangent vector $T$. By construction one takes
$S$ and $T$ commuting with each other. Also, if $\bar{S}$ is the 1-form 
associated to $S$ in such a way that 
\[\bar{S}(S) = 1 {\mbox{ then }}\bar{S}(T) = 0 \]
The closed 2-surface $\Sigma $ thus constructed has a north pole at ${\rm P}$ 
and a south pole at ${\rm O}$. We then  define ${\tilde{B}}$, the pre-dual 1-form 
functional associated to the dual 1-form functional 'connection' related to 
the dual theory, as
\begin{equation}
{\tilde{B}}({\rm P},{\Sigma }):=\int_{\Sigma}
G(\sigma ,t)\epsilon _{\mu \nu  \rho \lambda
}F^{\rho \lambda }(\sigma,t )T^{\mu }(\sigma,t )S_{\tau}(\sigma,t )S^{\tau}(\sigma,t )
¥d{X_{\sigma,t }^{\nu }}G^{-1}(\sigma ,t)d\sigma dt.
\label{6}¥\end{equation}
 
where $G(\sigma ,t)$ is given as
\[G(\sigma ,t) = G_{A} ( p(t),\sigma,\hat{\cal
{C}}_t) \equiv exp :\int_{ p(t)}^{\sigma}A_t : \]
$ p(t)$ is a generic point on the curve ${\cal {C}}$. In  
the language of forms the last equation may be written 
\begin{equation}
{\tilde{B}}(P,{\Sigma }):=\int_{\Sigma}
G(\sigma ,t) \star F \wedge \bar{S}(S,T) G^{-1}(\sigma ,t)d\sigma dt.
\end{equation}

Notice several interesting features of this definition. First, for the 
case of $U(1)$ abelian gauge theories eq.\ref{6} reduces to
\begin{equation}
{\tilde{B}}({\rm P},{\Sigma }):=\int_{\Sigma} \star F \wedge \bar{S}(S,T) d\sigma dt.
\end{equation}
which differs only on a factor with length dimension from
\[ \widetilde {\cal A}({\cal C}_{\;\rm{O,P}}) 
= \int_{\rm O}^{\rm P}  F(t') [T]  dt' \]
meaning that ${\tilde{B}}$ is equivalent to a flat 1-form functional 
representing, in path space, an ordinary $U(1)$ principal bundle. In this 
case, the bundle will be the dual bundle with Hodge -$\star$ curvature $\star F$ and 
connection $A*$.

Second, for the non abelian case, we could expand eq.\ref{6} to first 
order when we have close enough points ${\rm O}$ , ${\rm P}$ and 
small $\Sigma$, obtaining
  
\begin{equation}
{\tilde{B}}({\rm P},{\Sigma })\approx  (\star F_{o} + [F_{o}(\Sigma),\star 
F_{o}] + \ldots  ) (\triangle \sigma)^{2}
\end{equation}
this equation shows that even to first order it is necessary to 
introduce terms involving the commutator of the 2-form curvature and 
its Hodge -$\star$ dual to arrive to an approximation of a dual 
bundle. ${\tilde{B}}$, as we will show later, does not represent a 
non abelian principal bundle on $X$ since it does not in general 
solve the flat condition eq.\ref{7}. But, it will represent a 'twisted' 
version of the dual bundle since through a map involving an  automorphism of the Lie 
algebra and some other factors we get the non abelian dual in path 
space.

\section{ Yang-Mills field equations on path space} 

We will  only be interested in the application of $\tilde{B}(P,{\Sigma }%
) $ to deformations that are different from zero only
in a small neighborhood of the 2-surface ${\Sigma }$.

We show now that 
\begin{equation}
\widetilde {\Delta} {\tilde{B}}(P,{\Sigma })=0.
\end{equation}
over the Yang - Mills field equations.

We will denote ${ B_{\mu \nu} \equiv \epsilon_{\mu \nu \rho \lambda}
F^{\rho \lambda}}$ and compute the integral over the parameter $\rm t$ in 
terms of a discretization $(\rm t_i)$. Also, ${\hat{\cal{C}}}_{t_{i}}\equiv 
{\hat{\cal{C}}}_i$, and we  write the
arguments of $G$ only once in order to simplify the notation.

We have, 
\begin{eqnarray}
\widetilde {\Delta} {\tilde{B}} &=&\sum_{i=1}^{N}\oint_{{\hat{\cal{C}}}_i}[%
\widetilde{\Delta }G(\sigma ,{\hat{\cal{C}}}_i).G^{-1},G B_{\mu \omega }(\sigma
)T^{\mu }(\sigma )G^{-1}]\wedge d{X_{\sigma ,i}^{\omega }}d\sigma   \nonumber
\\
&&+\sum_{i=1}^{N}\oint_{{\hat{\cal{C}}}_i}G(\sigma ,{\hat{\cal{C}}}_i)(B_{\mu \omega
},\rho (\sigma )d{X_{\sigma ,i}^{\rho }}T^{\mu }(\sigma )  \nonumber \\
&&+B_{\mu \omega }T^{\mu },\rho d{X_{\sigma ,i}^{\rho }})G^{-1}\wedge d{%
X_{\sigma ,i}^{\omega }}d\sigma ,
\end{eqnarray}

\begin{equation}
\widetilde{\Delta }G(\sigma ,{\hat{\cal{C}}}_i).G^{-1}=-\widetilde{A}(\sigma ,%
{\hat{\cal{C}}}_i)+ G(\sigma ,{\hat{\cal{C}}}_i)A_{\rho }d{X_{\sigma ,i}^{\rho }}G^{-1},
\end{equation}

\begin{equation}
{T^{\mu }}_{,\rho }(\sigma )d{X_{\sigma ,i}^{\rho }}=(d{X_{\sigma ,i}^{\mu }}%
)^{\prime }
\end{equation}
where $^{\prime }$ denotes derivative with respect to $\sigma $, 
\[
B_{\mu \omega }{T^{\mu }}_{,\rho }d{X_{\sigma ,i}^{\rho }}\wedge d{X_{\sigma
,i}^{\omega }}=B_{\mu \omega }(d{X_{\sigma ,i}^{\mu }})^{\prime }\wedge d{%
X_{\sigma ,i}^{\omega }}=
\]
\begin{equation}
=\frac{1}{2}(B_{\mu \omega }(\sigma )d{X_{\sigma ,i}^{\mu }}\wedge d{%
X_{\sigma ,i}^{\omega }})^{\prime }-\frac{1}{2}B_{\mu \omega ,\rho }(\sigma
)T^{\rho }(\sigma )d{X_{\sigma ,i}^{\mu }}\wedge d{X_{\sigma ,i}^{\omega }}.
\end{equation}

After integration by parts, we obtain 
\begin{eqnarray}
\widetilde{\Delta}{\tilde{B}} &=&\sum_{i=1}^{N}\oint_{{\hat{\cal{C}}}_i}G(\sigma ,
{\hat{\cal{C}}}_i)\frac{1}{2}{\cal {D}}_{[\rho} B_{\mu \omega ]}(\sigma
)T^{\mu }(\sigma )G^{-1}d{X_{\sigma ,i}^{\rho }}\wedge d{X_{\sigma
,i}^{\omega }}d\sigma \Delta t  \nonumber \\
&&+\sum_{i=1}^{N}\oint_{{\hat{\cal{C}}}_i}[-\widetilde{A}(\sigma ,{\hat{\cal{C}}}_i), 
\nonumber \\
&&G(\sigma ,,{\hat{\cal{C}}}_i)B_{\mu \omega }(\sigma )T^{\mu }(\sigma )G^{-1}d{%
X_{\sigma ,i}^{\omega }}]d\sigma \Delta t  \nonumber \\
&&\sum_{i=1}^{N}\oint_{{\hat{\cal{C}}}_i}[\frac{1}{2}\widetilde{A}(\sigma ,{\hat{\cal{C}}}_i)(T),G(\sigma ,{\hat{\cal{C}}}_i)B_{\mu \omega }(\sigma )d{X_{\sigma ,i}^{\mu }%
}G^{-1}\wedge d{X_{\sigma ,i}^{\omega }}]d\sigma \Delta t
\nonumber \end{eqnarray}
the last two terms reduce to 
\[
\sum_{i=1}^{N} \oint_{{\hat{\cal{C}}}_i}[-G(t_{i},{\hat{\cal{C}}}_i)F_{\rho \lambda }(t_{i})G^{-1}S^{\rho
}(t_{i})d{X_{t_{i}}^{\lambda }},G(t_{i},{\hat{\cal{C}}}_i)B_{\mu \omega
}(t_{i})T^{\mu }(t_{i})G^{-1}d{X_{t_{i}}^{\omega }}] +
\]
\begin{equation}
+\frac{1}{2}\sum_{i=1}^{N} \oint_{{\hat{\cal{C}}}_i}[G(t_{i},{\hat{\cal{C}}}_i)F_{\rho \omega
}(t_{i})G^{-1}S^{\rho }(t_{i})T^{\lambda }(t_{i}),G(t_{i},{\hat{\cal{C}}}_i)
B_{\mu \omega }(t_{i})d{X_{t_{i}}^{\mu }}G^{-1}\wedge d{%
X_{t_{I}}^{\omega }}]
\end{equation}
when applied to deformations orthogonal to $S$, these equations will cancel.

We then have 
\begin{equation}
\widetilde{\Delta}{\tilde{B}}=\frac{1}{2}\sum_{i=1}^{N}\oint_{{\hat{\cal{C}}}_i}G(t_{i},{\hat{\cal{C}}}_i)
{\cal D}_{[\rho} B_{\mu \omega ]}(t_{i})T^{\mu }(t_{i})d{X_{t_{i}}^{\rho }}%
\wedge d{X_{t_{i}}^{\omega }}G^{-1}\Delta t
\end{equation}
which is zero over the Yang Mills field equations:
\begin{equation}
\widetilde {\Delta} {\tilde{B}}=0  	
	\label{4}
\end{equation}¥

Conversely, one may prove  by taking arbitary small neighborhoods of a point on $%
{\cal C}$,  and arbitrary deformations within these neighborhoods that 
one obtains the Yang-Mills field equations when
\begin{equation}
\widetilde {\Delta} {\tilde{B}}=0  \nonumber
\end{equation}

Further it is also possible to show see \cite{LIA}, that on a 
contractible open set $U$, $U \subset X$, where the curve ${\cal 
{C}}$ lies, for a Lie algebra valued 1-form functional $\widetilde{\lambda } 
(P, {\cal {C}})$ in a matrix representation satisfying
\begin{equation}
\widetilde{\Delta }\widetilde{\lambda }(P,{\cal C})=0
\end{equation}
there exists a $G'$, a path ordered exponential of a 1-form 
connection $B$, and a map $W$  such that 
\begin{equation}
\widetilde{\lambda }(P,{\cal C})=  W_{(P,{\cal C})}\cdot(\widetilde{\Delta }G'(P,{\cal C})\cdot 
G'^{-1}) 
\end{equation}
where $W_{(P,{\cal C})}$ is an  automorphism  of 
the Lie Algebra representable in terms of matrices.
If the structure group is $U(1)$  a solution for $W$ is readily found 
to be the identity map multipled by some factors $f_{D}$ that may be topological in 
nature or related to the different density degrees of the 1-form 
functionals linked by the map.  
\begin{equation}
W(P,{\cal C})={\rm 1\hspace{-0.9mm}l\hspace{-1mm}}\;f_{D}
\end{equation}
If instead $\widetilde {\lambda}$ is flat 
\begin{equation}
\widetilde{\Delta }\widetilde{\lambda }+\widetilde{\lambda }\wedge \widetilde{%
\lambda }=0,
\end{equation}
then as earlier stated, there exists a $G$ such that 
\begin{equation}
\widetilde{\lambda }=-\widetilde{\Delta }G\cdot G^{-1}=\widetilde{A}(P,{\cal C}).
\end{equation}

\section{ The dual map for $U(1)$ and $SU(N)$ bundles}

One may apply the last section results to get a solution for eq.\ref{4} in 
terms of a particular path ordered exponential $G*$ and a connection $A*$ 
with which one constructs the 1-form functional $\widetilde{\cal A}\star$ related 
to the dual theory,
\[ \widetilde {\cal A}\star({\cal C}_{\;\rm{O,P}})[\bullet] 
= \int_{\rm O}^{\rm P} G_{A\ast} ({\rm O},p(t'),{\cal C}) \star F(t') [T,\bullet] 
G_{A\ast} ({\rm
O},p(t'),{\cal C}^{-1}) dt' \] 

Thus, the Yang-Mills equations of motion are realized, 
via the pre-dual potential ${\tilde{B}}$ and the dual map $W$, by a flat 
1-form functional associated to a non abelian principal bundle on the 
manifold $X$ with curvature $\star F$ and connection $A*$. 

A possible solution for eq.\ref{4} would be, in a matrix representation  
 
\begin{equation}
{\tilde{B}}=  W_{(P,{\cal C},S)} ^{-1} (\widetilde{\Delta }G*(P,{\cal C})\cdot 
G*^{-1}) \cdot W_{(P,{\cal C},S)} f_{D}(S)\label{5}
\end{equation}.

The factor $f_{D}(S)$ has dimensions of length related to a particular 
radius of a loop in the family of loops $S_t$ (the closed deformations $S_t$ 
making up the surface $\Sigma$) .This factor, resembling a coupling 
constant, arises from the fact that the pre-dual potential ${\tilde{B}}$ is 
constructed from a surface and the ordinary potential in path space 
$\widetilde{A }$ from a curve.

In general,$ W_{(P,{\cal C},S)}$ will depend on the surface  $\Sigma$ 
and must satisfy $\widetilde{\Delta }W=0$ for any deformation of 
$\Sigma$. It will map to elements of the structure group and will have a structure of a 
path ordered exponential containing terms like
\[ exp (\int_{\Sigma} F(T,S)) \]
$ F(T,S)$ is the ordinary Lie Algebra valued 2-form curvature integrated 
along $T$ and $S$. This term that necessarily appears in $W$ introduces 
topological factors in the dual map that may be labelling the bundles.

We may understand the last assertions comparing  eq.\ref{6} and eq.\ref{5} 
in the limit of very close points ${\rm O}$ and ${\rm P}$ and small surfaces $\Sigma$.

\section{ Conclusions}

We have presented, in path space,  a complete set of equations 
representing the  Yang-Mills equations and Bianchi identities of 
ordinary non abelian gauge theories. Also we have introduced a new 
object the pre-dual 'potential' $\tilde{B(\rm {P},\cal{C},S}$ that allows 
the passage from a non abelian theory to its dual. We need further to 
study the universe of solutions for the dual map and its relation to 
classifying quantities for the unequivalent principal bundles.

{\it{Acknowledgements}}¥
I like to thank  A. Restuccia, L. Recht and E. Planchart for helpful discussions 
that made clearer the exposition of this article. Also, to Prof. 
M. Halpern for calling attention to several useful earlier references on the topic 
of this paper.

¥


\begin{thebibliography}{99}

\bibitem{Hal}M.B. Halpern, Phys.Rev. {\bf D19} (1979) 517; {\bf 
16}(1977) 1798.
\bibitem{Durand}L. Durand and E. Mendel, Phys.Rev. {\bf D26} (1982) 1368.
\bibitem{Kazama}Y. Kazama and R. Savit,  Phys.Rev. {\bf D21} (1980) 2916.
\bibitem{Alv}E. Verlinde, Nucl. Phys. {\bf B455} (1995) 211; Y.Lozano, Phys. Lett. {\bf
B364} (1995) 19 ;J.L.F. Barb\'on, Nucl. Phys. {\bf B452} (1995) 313; A. Kehagias, 
hep-th/9508159; J. Stephany, Phys. Lett.{\bf B 390} (1997) 128.
\bibitem{Alv2}M.Caicedo, I.Martin and A.Restuccia, Proceedings of 
1-SILAFAE , Mexico, Nov. 1996 (American Physical Society); hep-th/9701010. 
\bibitem{Bry}J.L.Brylinski, Prog. in Math. Vol.107  
$\underline{Loop\;Spaces,\;}$
$\underline{Characteristic\;Classes\; and \;
Geometric\; Quantization}$, $Birkh\ddot{a}user$, Boston
(1993),in here the extension is proven for curvature 3-forms 
associated to sheaf of grupoids. Also, it has recently been proven 
for the case of 4-forms by  Brylinski and McLaughlin (private communication).
\bibitem{Isb}I. Martin, "Non Abelian Dualities" . Proceedings of 1-SILAFAE , Mexico, Nov. 1996 
(American Physical Society). 
\bibitem{Poly}A.M.Polyakov, Nucl. Phys.{\bf B164}(1980) 171;  Nucl. 
Phys.{\bf B486}(1997) 23.
\bibitem{LIA}I.Martin and A.Restuccia in preparation.
\bibitem{Barre}J.W.Barrett,Int. J. Theo. Phys. {\bf 30}(1991) 1171 .  
\bibitem{Cho}C.Hong-Mo, J.Faridani and T. Sheung Tsun, Phys.Rev.{\bf 
D53}(1996) 7293. 
\bibitem{Bry2}J.L.Brylinski,"The Radon Transform and functionals on 
the space of curves". Gelfand Seminars Vol.2 , (1993).
\end{thebibliography}
\end{document}